\documentclass[journal=nalefd,manuscript=letter]{achemso}

\usepackage{amsmath,amssymb}
\usepackage{graphicx}
\usepackage{float}
\usepackage{bm}
\usepackage[colorlinks=true,citecolor=blue,urlcolor=blue,linkcolor=blue]{hyperref}

\title{Ferroelectric control of exciton photoluminescence in rhombohedrally stacked bilayer WSe$_2$}

\author{Wenlong~Zhong}
\affiliation{National Laboratory of Solid State Microstructures and Department of Physics, Nanjing University, Nanjing 210093, China}

\author{Kenji~Watanabe}
\affiliation{Research Center for Electronic and Optical Materials, National Institute for Materials Science, Tsukuba 305-0044, Japan}

\author{Takashi~Taniguchi}
\affiliation{Research Center for Materials Nanoarchitectonics, National Institute for Materials Science, Tsukuba 305-0044, Japan}

\author{Xiaoxiang~Xi}
\email{xxi@nju.edu.cn}
\affiliation{National Laboratory of Solid State Microstructures and Department of Physics, Nanjing University, Nanjing 210093, China}
\alsoaffiliation{Collaborative Innovation Center of Advanced Microstructures, Nanjing University, Nanjing 210093, China}
\alsoaffiliation{Jiangsu Physical Science Research Center, Nanjing 210093, China}

\begin{document}

\maketitle
\noindent
\textbf{Abstract}\\
\textbf{Sliding ferroelectricity offers a non-volatile route to electrically programmable electronic and optical functionality, but deterministic control of excitonic photoluminescence is often obscured by moir\'{e} reconstruction and domain averaging. Here we study an artificially assembled WSe$_2$ bilayer with near-zero twist in rhombohedral stacking and observe reversible, hysteretic electrical switching through excitonic photoluminescence. Under an out-of-plane electric field, momentum-indirect QK exciton emission exhibits Stark-slope reversals and abrupt energy jumps, resolving a four-stage sequence separated by two band realignments and one ferroelectric switching event. Polarization reversal also selects the momentum-direct KK interlayer-exciton branch that hybridizes with an intralayer exciton, producing history-dependent photoluminescence. Ferroelectric modulation of resonant Raman scattering provides a complementary readout, with hysteretic intensities correlated with the polarization-dependent interlayer field. These results establish an exciton-based quantification of the electronic structure in ferroelectric WSe$_2$ and open a route to electrically programmable optoelectronic functionality.}

\maketitle

\vspace{3mm}
\noindent
\textbf{Introduction}\\
Sliding ferroelectricity has emerged as a new frontier for ultrathin ferroelectrics and, more broadly, ordered states in the two-dimensional (2D) limit~\cite{Fei2018,Wu2021,Vizner2021,Yasuda2021,Wan2022,Wang2022,Weston2022}. Unlike conventional 2D ferroelectrics based on polar unit cells and ionic displacements~\cite{Chang2016,Liu2016,Xiao2018}, this form of ferroelectricity arises from interlayer charge transfer enabled by inversion-asymmetric van der Waals stacking~\cite{Wu2021}. An out-of-plane electric field can drive in-plane domain-wall motion, switching between bistable stacking configurations and reversing the polarization~\cite{Ke2025}. Enabled by clean van der Waals interfaces and low switching barriers, sliding ferroelectrics offer a route to ultrafast, low-power, high-endurance functionality that can be programmed electrically and retained without bias~\cite{Wang2021,Li2024,Wu2023,Bian2024,Yasuda2024}.

A particularly appealing aspect of sliding ferroelectricity is its ability to couple to, and thereby control, intrinsic and emergent properties of the host system. Experimentally, sliding ferroelectricity has enabled switching of topological charge transport~\cite{Chen2024}, Berry-curvature-related responses~\cite{Xiao2020}, and superconductivity~\cite{Jindal2023,Klein2023}. Extending such non-volatile control to optical responses is attractive because optical readout is contactless, spatially resolved, and energy-resolved. Semiconducting transition metal dichalcogenides provide a natural platform, hosting tightly bound excitons and valley-selective optical selection rules in the monolayer limit~\cite{Mak2010,Splendiani2010,Zhao2013,Mak2016,Wang2018RMP}. In bilayers, interlayer coupling reshapes band alignment and exciton wavefunctions, enabling electric-field tuning of Stark shifts and interlayer–intralayer hybridization~\cite{Rivera2015,Wang2018,Peimyoo2021,Huang2022,Feng2024}. Photoluminescence (PL) provides a bright, energy-resolved signal, making it well suited for quantitative optical readout of excitonic states. 

Despite recent progress, ferroelectric control of exciton PL remains challenging to establish cleanly. In marginally twisted homobilayers, lattice reconstruction often produces alternating AB and BA domains within a diffraction-limited spot, complicating spectral assignment and obscuring deterministic switching of a specific emissive state~\cite{Sung2020,Andersen2021}. Coupling between ferroelectric order and excitonic spectra has been reported in 3R-MoS$_2$~\cite{Liang2022,Yang2024,Deb2024,Liang2025PRX,Liang2025,Ouyang2025} and in rhombohedrally stacked (R-stacked) MoSe$_2$ and WSe$_2$ bilayers~\cite{Sung2020,Li2026}, providing key insight into switching pathways and band parameters. However, resolving how polarization switching and field-induced band realignment jointly shape excitonic PL remains challenging, particularly when the two processes occur on comparable field scales.

In this work, we use an artificially assembled WSe$_2$ bilayer with near-zero twist in R stacking to investigate reversible, hysteretic electrical switching and its coupling to intralayer and interlayer excitons using spatially resolved PL measurements. The momentum-indirect QK-exciton PL resolves a four-stage Stark shift comprising two field-driven valence-band realignments and one ferroelectric switching event. Ferroelectric switching also selects which KK interlayer-exciton branch hybridizes with the lower-energy intralayer exciton. Hysteretic Raman intensities approximately follow a common effective-field dependence for each mode, providing a complementary readout through ferroelectric modulation of resonant Raman scattering. Quantitative analysis yields key band parameters, and these results open a route to non-volatile optoelectronic functionality based on ferroelectrically programmable excitons.

\begin{figure}[t]
\centering
\includegraphics[width=0.65\linewidth]{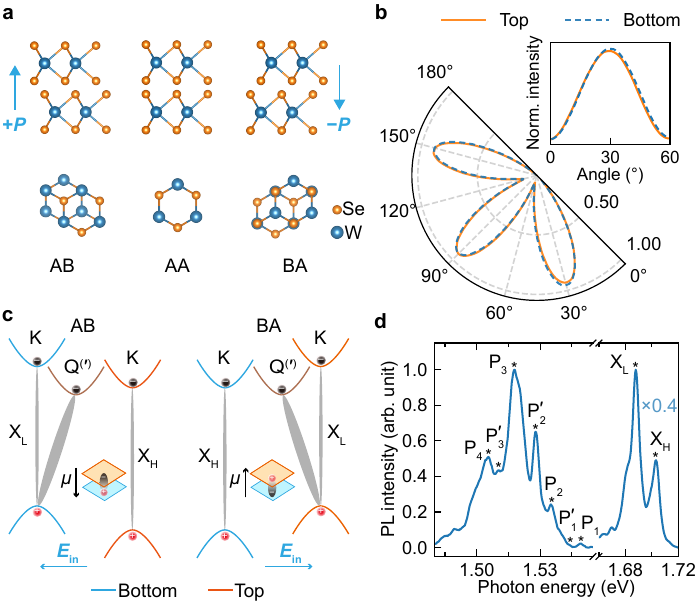}
\caption{\textbf{Stacking configuration, band alignment, and excitonic PL in R-stacked bilayer WSe$_2$.} \textbf{a} Side and top views of AB, AA, and BA stacking for bilayer WSe$_2$. $\pm P$ denotes the ferroelectric polarization. \textbf{b} Polarization-resolved SHG from the top and bottom monolayers at 300~K. \textbf{c} Schematic band alignment for AB and BA stacking, highlighting momentum-direct A excitons (X$_{\mathrm{L}}$ and X$_{\mathrm{H}}$) and momentum-indirect QK (or Q$'$K) excitons. The built-in field $E_{\text{in}}$ and interlayer-exciton dipole moment $\mu$ are indicated. \textbf{d} PL spectra of R-stacked bilayer WSe$_2$ at 5~K, with interlayer and intralayer exciton peaks labeled.}
\label{Fig1}
\end{figure}

\vspace{2mm}
\noindent
\textbf{Results and Discussion}\\
Near-parallel stacking of two WSe$_2$ monolayers produces a moir\'{e} superlattice with three high-symmetry local registries: AB (W above Se), BA (Se above W), and AA (vertical alignment)~\cite{Li2017,Naik2018}. AB and BA are inversion-asymmetric and related by mirror reflection about the horizontal plane (Figure~\ref{Fig1}a). For AB stacking, interlayer charge transfer from the top to the bottom layer generates an out-of-plane polarization pointing upward~\cite{Li2017}, defined as the positive direction throughout this work. We focus on a near-zero-twist bilayer sample with micron-scale domains suitable for far-field optical microscopy. Polarization-resolved second-harmonic generation (SHG)~\cite{Seyler2015} measured in the non-overlapping regions of the constituent layers (Figure~\ref{Fig1}b) yields a relative angular offset of approximately $0.5^\circ$ between these reference regions (Figure~S10 and Methods). All data in the main text were acquired from this sample, and the sample temperature was 5~K unless specified otherwise. At larger twist angles, lattice reconstruction can produce mixed AB and BA microdomains that complicate quantitative analysis~\cite{Sung2020} (see data for an approximately 1$^{\circ}$ twisted sample in Section~S2).

Exciton spectroscopy provides a sensitive probe of sliding ferroelectricity, because the Stark effect directly reflects the built-in field $E_{\mathrm{in}}$ and the associated type-II band alignment~\cite{Liang2022}. In AB stacking, a downward $E_{\mathrm{in}}$ implies a higher electrostatic potential in the top layer and shifts its bands to lower energy, while BA stacking shows the opposite shift (Figure~\ref{Fig1}c). Interlayer hybridization lowers the Q and Q$'$ valleys~\cite{Scuri2020}, favoring momentum-indirect QK and Q$'$K interlayer excitons. Their phonon-assisted recombination~\cite{Wang2018,Li2022,Altaiary2022,Huang2022} yields a replica series near 1.52~eV (Figure~\ref{Fig1}d). Following Ref.~\cite{Li2022}, we label the QK replicas P$_1$--P$_4$ and the Q$'$K replicas P$_1'$--P$_3'$, and refer to them collectively as QK excitons for brevity. The peak energies are consistent with prior reports~\cite{Scuri2020,Li2022}, with a small offset of $\sim$5.3~meV likely due to unintentional doping or local strain introduced during device fabrication (Section~S3). The intrinsic layer inequivalence associated with inversion-symmetry breaking in R stacking also lifts the degeneracy of the intralayer K-valley A exciton~\cite{Li2022,Liang2022}, producing two peaks near 1.70~eV (Figure~\ref{Fig1}d). In AB (BA) stacking, the bottom-layer intralayer exciton has lower (higher) energy, denoted X$_{\mathrm{L}}$ (X$_{\mathrm{H}}$), as illustrated in Figure~\ref{Fig1}c.

\begin{figure}[t]
\centering
\includegraphics[width=0.65\linewidth]{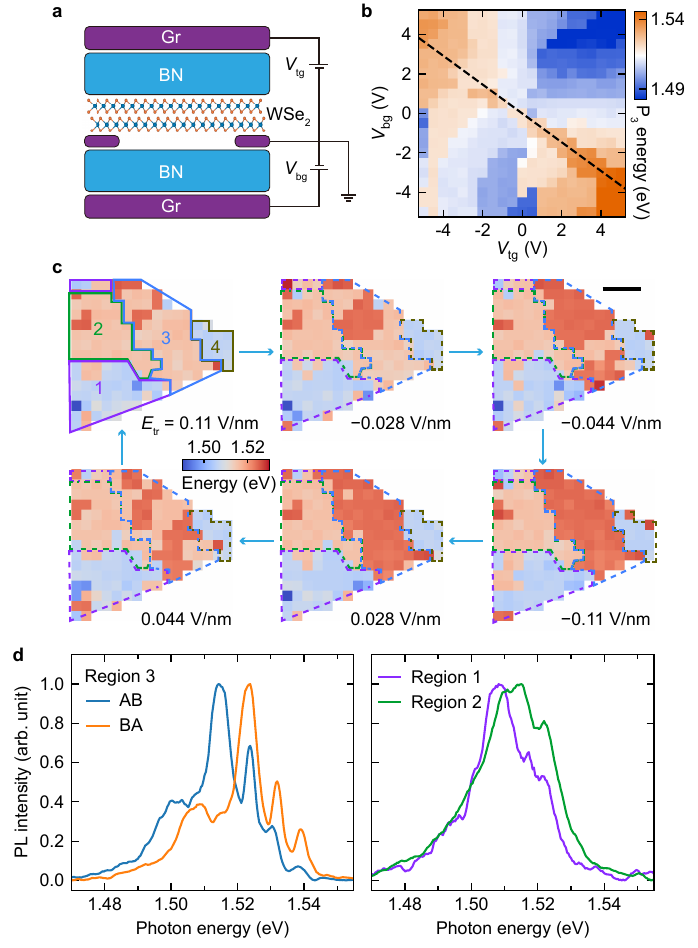}
\caption{\textbf{PL mapping of ferroelectric switching.} \textbf{a} Schematic of the dual-gate device. \textbf{b} Interlayer exciton P$_3$ energy as a function of top- and back-gate voltages, with the charge-neutrality line marked. \textbf{c} Spatial maps of the dominant interlayer-exciton PL energy (P$_3$ when resolved, otherwise the spectral maximum between 1.49--1.53~eV), acquired after applying a sequence of training fields $E_{\mathrm{tr}}$. Scale bar: 3~$\mu$m. Different regions are outlined and labeled. \textbf{d} PL spectra measured from AB- and BA-stacked domains in Region 3 (left) and from Regions 1 and 2 (right). Data in \textbf{c} and \textbf{d} were acquired at $E_{\mathrm{ex}}=-0.011~\mathrm{V/nm}$. }
\label{Fig2}
\end{figure}

First-principles calculations show that the Q-valley conduction-band states are delocalized across the two layers, whereas the K-valley valence-band states are strongly layer polarized~\cite{Ferreira2021}. Consequently, the QK exciton possesses an out-of-plane dipole component $\mu$, with $\mu<0$ ($\mu>0$) for AB (BA) stacking (Figure~\ref{Fig1}c). We tuned the excitons using a dual-gated device (Figure~\ref{Fig2}a). Sweeping the top- and bottom-gate voltages yielded a map of the dominant P$_3$ peak energy (Figure~\ref{Fig2}b). Along the charge-neutrality line, the P$_3$ energy varies linearly with the gate-induced external field $E_{\text{ex}}$, showing a pronounced quantum-confined Stark shift (Section~S3), while perpendicular tuning softens the peak, consistent with doping-induced renormalization and possible exciton-polaron formation~\cite{Li2026}. In the following, we focus on the charge-neutrality line and study the excitonic PL as a function of $E_{\text{ex}}$.

For weak fields well below the coercive field, the Stark shift $\Delta \epsilon=-\mu E_{\mathrm{ex}}$ provides a direct readout of the polarization state~\cite{Yang2024}. Figure~2d compares interlayer-exciton PL spectra of the two polarization states measured at $E_{\mathrm{ex}}=-0.011$~V/nm. For AB stacking ($\mu<0$), the negative field yields $\Delta \epsilon<0$ and shifts the emission to lower photon energy. The complementary spectrum is therefore assigned to BA stacking. The two states are cleanly resolved by a $\sim$11~meV splitting of the entire replica series. We then imaged domains by mapping the dominant interlayer-exciton PL energy (P$_3$ when resolved; otherwise the spectral maximum between 1.49--1.53~eV) at a fixed probe field after training with $E_{\mathrm{tr}}$ (Figure~\ref{Fig2}c). Regions with distinct responses emerge. Region~1 overlaps a graphite electrode and Region~4 is controlled by a single gate, and in both cases screening by the electrode or gate-induced doping suppresses ferroelectric switching. Regions~2 and 3 lie between the two gates, but an optical switching signal is detected only in Region~3 within the investigated electric-field range (Section~S6). Varying $E_{\mathrm{tr}}$ reveals progressive growth of the favored polarization domains. The resulting domain patterns remain unchanged during PL mapping at the probe field of $E_{\mathrm{ex}}=-0.011~\mathrm{V/nm}$.

To probe local switching dynamics, we fixed the laser spot in Region 3 and performed field-dependent PL spectroscopy. Figures~3a and 3c show the forward and backward sweeps, respectively. All QK exciton peaks exhibit piecewise-linear Stark shifts interrupted by abrupt jumps at coercive fields, which mark ferroelectric switching. In the forward sweep, the coercive field was $E_{\text{c}}^{\text{f}}\approx 0.033$~V/nm. In the backward sweep, two jumps occurred at $E_{\text{c1}}^{\text{b}}\approx -0.019$~V/nm and $E_{\text{c2}}^{\text{b}}\approx -0.033$~V/nm, indicating two domains within the laser spot with distinct coercive fields, likely due to extrinsic pinning. The faint continuations of the PL branches in Figures~3a and 3c may originate from small pinned domains within the optical spot. These domains may retain their stacking configuration over the measured field range while contributing residual Stark-shifted emission. Measurements at another position in the same device and in a second device reproduce the characteristic switching-related spectral response, with variations in the switching thresholds (Section~S8, Figures~S11 and S12).

\begin{figure*}[t]
\centering
\includegraphics[width=0.95\linewidth]{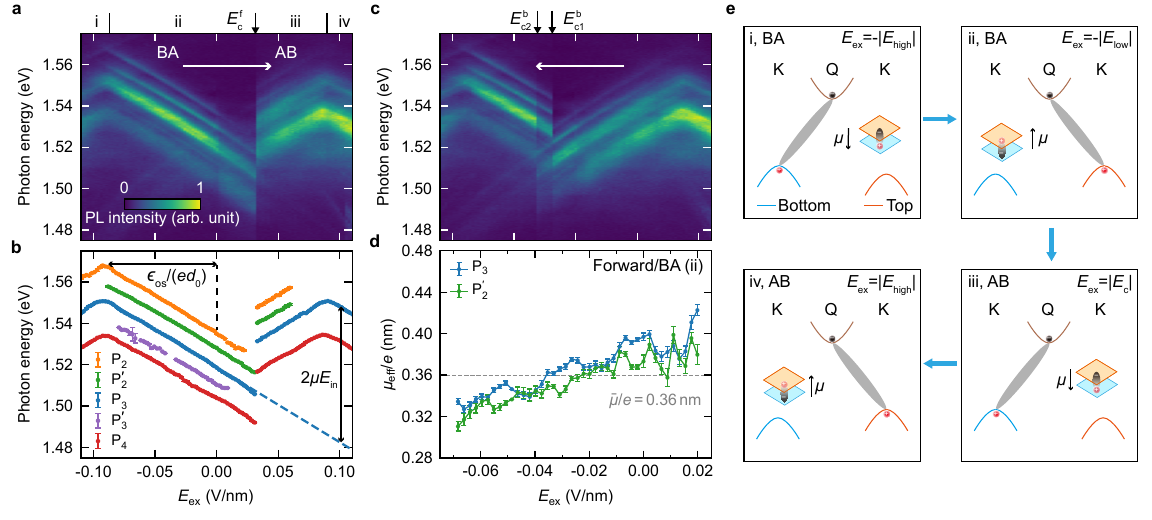}
\caption{\textbf{Ferroelectric control of interlayer-exciton PL.} \textbf{a,c} Field-dependent QK exciton PL spectra acquired during the forward (\textbf{a}) and backward (\textbf{c}) sweeps. Horizontal arrows indicate the sweep direction, and vertical arrows mark the coercive fields. \textbf{b} Peak energies extracted from \textbf{a}. Dashed lines and arrows label the energy offset from the built-in field ($2\mu E_{\text{in}}$) and the valence-band offset ($\epsilon_{\text{os}}$). Error bars are obtained from spectral fitting. \textbf{d} Effective dipoles $\mu_{\mathrm{eff}}/e$ of $P_3$ and $P_2^{\prime}$ in the forward-sweep BA interval (ii), extracted from adjacent-point Stark slopes using a 10-point moving average. Error bars denote propagated standard errors. The dashed line marks the average linear-Stark reference $\bar{\mu}/e=0.36$~nm. \textbf{e} Schematic evolution of the band alignment, the QK exciton, and its dipole moment during the forward sweep across the four field ranges i--iv indicated in \textbf{a}.}
\label{Fig3}
\end{figure*}

Remarkably, the Stark-slope sign changes three times in both sweeps. This reflects successive reversals of $\mu$ driven by field-induced band realignment and by ferroelectric switching, as schematized in Figure~\ref{Fig3}e for the forward sweep. At small $E_{\text{ex}}$ (Figure~\ref{Fig3}e-ii), the probed area is in BA stacking, inferred from the negative Stark slope near zero field. Negative $E_{\text{ex}}$ increases the top-layer potential and shifts its bands downward. At sufficiently large magnitude, this overrides the intrinsic band offset and relocates the valence-band maximum to the bottom layer (Figure~\ref{Fig3}e-i), yielding $\mu<0$ and a positive Stark slope. Positive $E_{\text{ex}}$ instead raises the top-layer bands, extending the near-zero-field trend until $E_{\text{c}}^{\text{f}}$ is reached, where the stacking switches to AB. This reverses the intrinsic offset (Figure~\ref{Fig3}e-iii) and produces a sudden Stark-slope change. With further increase of $E_{\text{ex}}$, the valence-band maximum relocates again (Figure~\ref{Fig3}e-iv), flipping the signs of $\mu$ and the Stark slope once more.

Figure~3b shows the fitted peak energies for the forward sweep. The average dipole $\bar{\mu}$ extracted from the approximately linear phonon-replica branches corresponds to an effective electron-hole separation of about 0.36~nm, approximately 55\% of the interlayer spacing $d_0\approx 0.65$~nm~\cite{Wang2022}, consistent with delocalized Q-valley electrons. Figure~3d shows the field dependence of the effective dipole, $\mu_{\mathrm{eff}}=-\mathrm{d}\epsilon/\mathrm{d}E_{\mathrm{ex}}$, for the representative P$_3$ and P$_2^{\prime}$ replicas in the pre-switching BA interval. Both exhibit an overall increase with $E_{\mathrm{ex}}$, consistent with continuous redistribution of Q-valley electron weight while the K-valley hole remains in the top layer (Section~S1). The parallel branches correspond to opposite polarization states, and their energy offset yields the built-in field via $\Delta \epsilon=2\mu E_{\mathrm{in}}$. From $\Delta \epsilon\approx 64.7$~meV, we obtain $E_{\mathrm{in}}\approx 0.091$~V/nm, corresponding to an interlayer potential difference of 59.2~mV. This value agrees with estimates from transport and Kelvin probe microscopy~\cite{Wang2022,Deb2022} and from calculations~\cite{Ferreira2021} (see Supporting Information for a quantitative comparison). The field $E_{\text{ex}}^{\text{i-ii}}$ required to relocate the valence-band maximum between the two layers, corresponding to the boundary between the field ranges (i) and (ii) in Figure~\ref{Fig3}a, provides an estimate of the K-valley valence-band offset, $\epsilon_{\text{os}}=ed_0|E_{\text{ex}}^{\text{i-ii}}|\approx 58.5$~meV. This value is close to the interlayer potential difference analyzed above.

\begin{figure*}[t]
\centering
\includegraphics[width=0.85\linewidth]{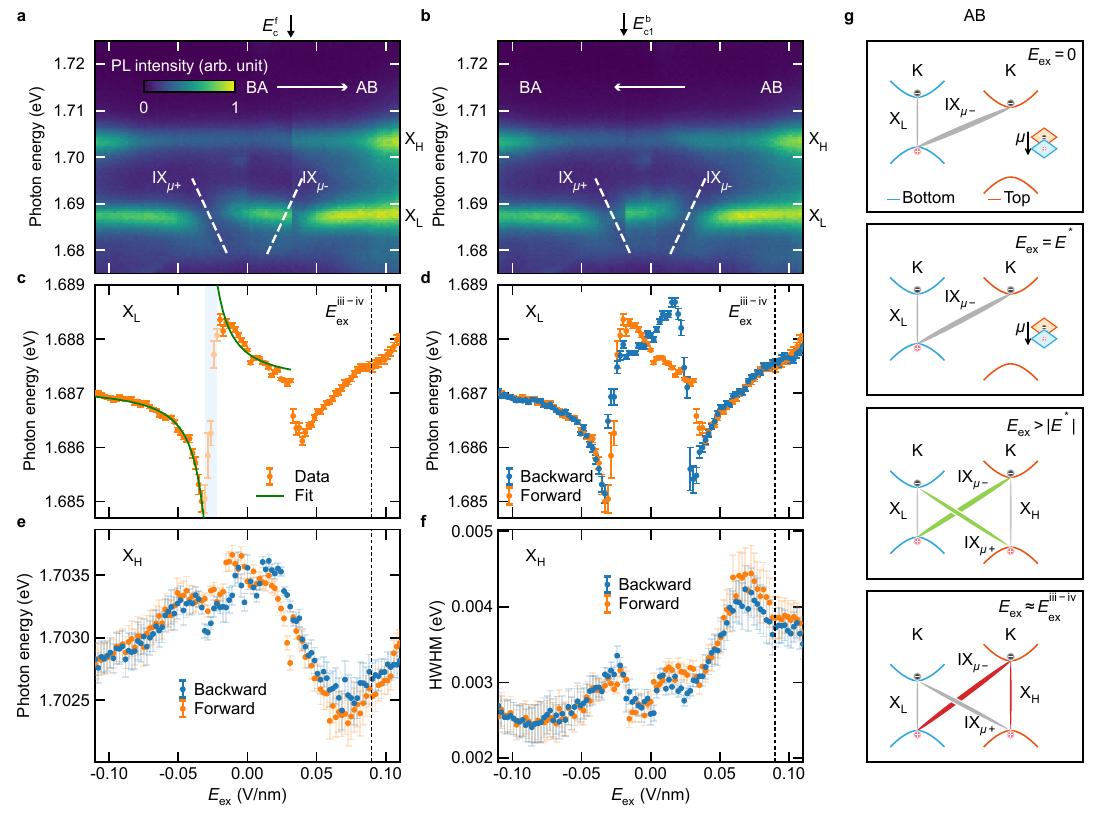}
\caption{\textbf{Ferroelectric switching of interlayer-intralayer exciton hybridization.} \textbf{a},\textbf{b} Field-dependent intralayer-exciton PL acquired in the forward (\textbf{a}) and backward (\textbf{b})  sweeps. Dashed lines mark avoided crossings arising from hybridization between the intralayer exciton X$_{\mathrm{L}}$ and the interlayer excitons IX$_{\mu+}$ or IX$_{\mu-}$. \textbf{c}--\textbf{f} X$_{\mathrm{L}}$ energy (\textbf{c},\textbf{d}), X$_{\mathrm{H}}$ energy (\textbf{e}) and half-width at half-maximum (\textbf{f}) extracted from \textbf{a} and \textbf{b}. Solid lines in \textbf{c} are fits to the model described in the text. Data in the shaded region are excluded from the fits.   Error bars in \textbf{c}--\textbf{f} are standard deviations obtained from fitting analysis. \textbf{g} Schematics of the band alignment and the relevant excitons at representative $E_{\text{ex}}$.}
\label{Fig4}
\end{figure*}

Interlayer excitons can also be momentum-direct at the K valley. We resolve this species when its Stark shift tunes it into resonance with the intralayer X$_{\text{L}}$ exciton, leading to hybridization (Figure~\ref{Fig4}). In AB stacking at zero field, the KK interlayer exciton comprises an electron in the top layer and a hole in the bottom layer, yielding $\mu<0$ and denoted IX$_{\mu-}$ (Figure~\ref{Fig4}g, top panel). Starting from AB stacking, positive $E_{\text{ex}}$ raises the top-layer bands and increases the IX$_{\mu-}$ energy (Figure~\ref{Fig4}g, second panel). When its detuning from X$_{\mathrm{L}}$ is small, the two excitonic states hybridize, producing the avoided crossing marked by the right dashed line in Figure~\ref{Fig4}b. In the backward sweep, both X$_{\mathrm{H}}$ and X$_{\mathrm{L}}$ show ferroelectric switching at $E_{\text{c1}}^{\text{b}}$, after which IX$_{\mu+}$ becomes relevant and produces an avoided crossing with the opposite Stark-shift slope (Figure~\ref{Fig4}b, left dashed line). The forward-sweep data in Figure~\ref{Fig4}a show analogous behavior.

Figures~\ref{Fig4}c and~\ref{Fig4}d show the extracted X$_{\mathrm{L}}$ energy (Section~S4), which exhibits clear hysteresis associated with ferroelectric switching. We model the hybridization using an effective two-level Hamiltonian,
\begin{equation}
H(E_{\text{ex}}) =
\begin{pmatrix}
\epsilon_{\text{X}_{\text{L}}} & g \\
g & \epsilon_{\text{IX}}^{0} - \mu E_{\text{ex}}
\end{pmatrix},
\end{equation}
which yields the energies of two hybridized branches as a function of $E_{\text{ex}}$, the coupling strength $g$, and the zero-field detuning  $\delta_0=\epsilon_{\mathrm{IX}}^{0}-\epsilon_{\mathrm{X}_{\mathrm{L}}}$ between the interlayer and intralayer exciton energies. $\epsilon_{\mathrm{X}_{\mathrm{L}}}$ is taken to be field independent because X$_{\mathrm{L}}$ is intralayer. The field dependence in Figures~\ref{Fig4}c and~\ref{Fig4}d is well described by this model  (Section~S4). For BA, the fits yield $g=3.4\pm0.5~\mathrm{meV}$, $|\mu|=0.64\pm0.21~e\cdot\mathrm{nm}$, and $\delta_0=-17.4\pm5.7~\mathrm{meV}$. For AB, the corresponding values are $g=3.3\pm0.5~\mathrm{meV}$, $|\mu|=0.64\pm0.20~e\cdot\mathrm{nm}$, and $\delta_0=-17.2\pm5.4~\mathrm{meV}$. The fitted dipole moments are consistent with layer-polarized electrons and holes at the K valley spatially separated by $d_0$. The analysis also yields the characteristic field $|E^*|=|\delta_0/\mu|=0.027$~V/nm for the strongest hybridization, consistent with the experimental observation.

The extracted coupling strength $g=3.3$~meV is much smaller than the other energy scales discussed above, consistent with symmetry constraints in R-stacked transition metal dichalcogenides. Specifically, $C_{3}$ symmetry suppresses direct interlayer hopping between K-valley conduction-band states of the two layers due to a mismatch in total azimuthal quantum number~\cite{Liang2022,Tong2017}. Such interlayer mixing is nonetheless required to hybridize interlayer and intralayer excitons. The finite but weak coupling observed here likely arises from symmetry-breaking or higher-order processes, such as phonon-assisted coupling. Similar hybridization has been reported in 3R-MoS$_2$~\cite{Liang2025} and R-stacked MoSe$_2$~\cite{Sung2020}. Here we show that it can be reversibly controlled by ferroelectric switching.

For $E_{\text{ex}}>E^*$ in AB stacking, the band alignment is expected to evolve until $\epsilon_{\mathrm{IX}_{\mu+}}=\epsilon_{\mathrm{IX}_{\mu-}}$, with the interlayer-exciton energy lying between $\epsilon_{\mathrm{X}_{\mathrm{L}}}$ and $\epsilon_{\mathrm{X}_{\mathrm{H}}}$ (Figure~\ref{Fig4}g, third panel). At higher fields, near $E_{\text{ex}}^{\text{iii-iv}}$ determined from Figure~\ref{Fig3}a, which separates field ranges (iii) and (iv), the layer-resolved K-valence bands become aligned (Figure~\ref{Fig4}g, bottom panel). In principle, this could lead to hybridization of $\mathrm{X}_{\mathrm{L}}$ ($\mathrm{X}_{\mathrm{H}}$) with IX$_{\mu+}$ (IX$_{\mu-}$). No clear evidence of such hybridization is observed for $\mathrm{X}_{\mathrm{L}}$, whereas the enhanced $\mathrm{X}_{\mathrm{H}}$ intensity at high fields and anomalies in its energy and linewidth near $E_{\text{ex}}^{\text{iii-iv}}$ suggest possible hybridization. The difference between the field at which these features occur and $E_{\text{ex}}^{\text{iii-iv}}$ may reflect different binding energies of QK interlayer and intralayer excitons. A plausible reason for the weak hybridization at high fields is that such hybridization would require sharing K-valley conduction-band electrons, but electrons preferentially occupy the lower-energy Q valley, stabilizing QK excitons and suppressing the K-K hybridization channel. Within this simplified picture, a similar high-field response would be expected on the positive and negative field sides, in contrast to the asymmetry observed in Figures~\ref{Fig4}e and \ref{Fig4}f. Further measurements on samples with improved quality are needed to clarify the high-field intralayer-exciton response.

\begin{figure*}[t]
\centering
\includegraphics[width=\linewidth]{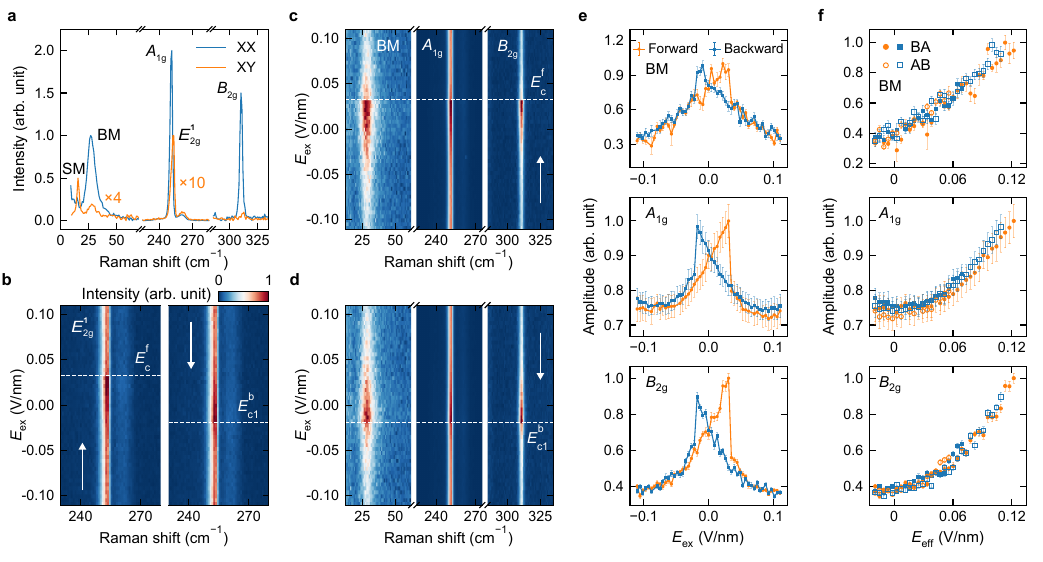}
\caption{\textbf{Ferroelectric modulation of resonant Raman scattering.} \textbf{a} Raman spectra of R-stacked bilayer WSe$_2$ at zero field measured in parallel (XX) and crossed (XY) polarization configurations. \textbf{b} XY-channel Raman intensity maps versus $E_{\text{ex}}$. \textbf{c,d} Corresponding XX-channel maps. Arrows in \textbf{b}--\textbf{d} indicate the sweep directions. \textbf{e} Field dependence of the fitted peak amplitude for the breathing mode (BM), $A_\mathrm{1g}$ mode, and $B_\mathrm{2g}$ mode. Error bars are standard errors obtained from fitting analysis. \textbf{f} Peak amplitudes from \textbf{e} replotted as a function of the effective interlayer electric field $E_{\mathrm{eff}}$.}
\label{Fig5}
\end{figure*}

We further reveal ferroelectric switching of excitonic states through resonantly enhanced Raman scattering. With 532~nm excitation, the dominant first-order Raman modes include the breathing (BM), $A_\mathrm{1g}$, and $B_\mathrm{2g}$ modes in the parallel polarization (XX) channel, and the shear (SM) and $E_\mathrm{2g}^{1}$ modes in the crossed polarization (XY) channel (Figure~\ref{Fig5}a)~\cite{Lin2021,Li2022,Pan2022}. Figures~5b--d plot field-dependent Raman intensity maps for representative modes during forward and backward sweeps. All modes show hysteretic intensity changes, most pronounced in the XX channel, with coercive fields matching those extracted from PL. The analyzed peak intensities show hysteretic butterfly-shaped loops (Figure~\ref{Fig5}e), while the mode frequencies and linewidths remain essentially unchanged (Section~S5). To relate this hysteresis to the interlayer electrostatics, we replot the amplitudes against $E_{\mathrm{eff}}=E_{\mathrm{in}}+sE_{\mathrm{ex}}$, where $s=+1$ for BA and $s=-1$ for AB. The built-in field $E_{\mathrm{in}}=0.091$~V/nm is fixed by the QK-exciton analysis (Figure~5f and Section~S5). For each phonon mode, the sweep segments collapse onto a common effective-field dependence, consistent with modulation of the Raman resonance condition by the interlayer electrostatics.

This behavior indicates field-tunable resonance rather than phonon renormalization. Raman intensities are enhanced when the incident or scattered photon is resonant with an intermediate electronic state~\cite{Yu2010}. In R-stacked bilayer WSe$_2$, 532~nm excitation is resonant with the C (also called A$'$) exciton~\cite{McCreary2022}. Ferroelectric switching shifts the relevant excitonic resonance through band-alignment changes, producing the observed hysteretic Raman intensity. The resonant enhancement weakens at higher temperature due to thermal broadening (Section~S5). The strongest modulation occurs for XX-channel modes dominated by out-of-plane vibrations~\cite{Pan2022}, consistent with their stronger influence on interlayer coupling and charge transfer  that underpin the ferroelectricity.

It is also worth noting that the hierarchy of band realignments inferred from the interlayer-exciton Stark response was not well resolved in bilayer 3R-MoS$_2$~\cite{Yang2024} or in R-stacked MoSe$_2$~\cite{Sung2020}. In MoS$_2$, ferroelectric switching was observed, but the Stark-slope sign appears nearly unchanged across switching~\cite{Yang2024}. This is due to the coercive field being close to the field required to reverse the relative band alignment, which renders the window corresponding to Regime~iii in Figure~\ref{Fig3}a vanishingly small. In MoSe$_2$, ferroelectric switching was not observed~\cite{Sung2020}, likely because the coercive field is large. Across MoSe$_2$, MoS$_2$, and WSe$_2$, the field required for interlayer-exciton dipole reversal lies in the range 0.06--0.09~V/nm, whereas the coercive field is much smaller in WSe$_2$ ($\sim$0.03~V/nm) than in MoS$_2$ ($\sim$0.07~V/nm)~\cite{Sung2020,Yang2024}. The separation of field scales in WSe$_2$ allows the full hierarchy of band realignments driven by both band shifts and ferroelectric switching to be clearly resolved. The reduced coercive field is favorable for low-threshold operation. An important next step is to identify the factors that control the coercive field and its variability. The exciton-based parameters extracted here provide quantitative inputs for engineering ferroelectrically programmable optical functionality in WSe$_2$ devices.

\vspace{2mm}
\noindent
\textbf{Data Availability}\\
All data are presented in the main text and Supporting Information.

\vspace{2mm}
\noindent
\textbf{Supporting Information}\\
Sample and device preparation; optical spectroscopy measurements; determination of twist angle and exciton dipole moment; data for an approximately 1° twisted bilayer sample; analysis of momentum-indirect interlayer excitons and interlayer–intralayer exciton hybridization; analysis of Raman-active phonon modes; data for Region 2; additional device characteristics; additional measurements of ferroelectric switching; comparison of the built-in interlayer potential with previous reports (PDF).

\vspace{2mm}
\noindent
\textbf{Author Contributions}\\
X.X. conceived the project. W.Z. performed the experiments. K.W. and T.T. grew the hBN crystals. W.Z. and X.X. analysed the experimental data, interpreted the results, and wrote the paper, with comments from all authors.

\vspace{2mm}
\noindent
\textbf{Competing Financial Interest}\\
The authors declare no competing financial interest.

\vspace{2mm}
\noindent
\textbf{Acknowledgments}\\
X.X. acknowledges support from the National Key Research and Development Program of China (Grant No.~2024YFA1409100), the Natural Science Foundation of Jiangsu Province (Grant Nos. BK20231529 and BK20233001), and the National Natural Science Foundation of China (Grant Nos. U24A6002 and 12474170), and the Fundamental Research Funds for the Central Universities (Grant No.~0204-14380264). K.W. and T.T. acknowledge support from the CREST (JPMJCR24A5), JST and World Premier International Research Center Initiative (WPI), MEXT, Japan.

\providecommand{\latin}[1]{#1}
\makeatletter
\providecommand{\doi}
  {\begingroup\let\do\@makeother\dospecials
  \catcode`\{=1 \catcode`\}=2 \doi@aux}
\providecommand{\doi@aux}[1]{\endgroup\texttt{#1}}
\makeatother
\providecommand*\mcitethebibliography{\thebibliography}
\csname @ifundefined\endcsname{endmcitethebibliography}
  {\let\endmcitethebibliography\endthebibliography}{}

\begin{mcitethebibliography}{53}
\providecommand*\natexlab[1]{#1}
\providecommand*\mciteSetBstSublistMode[1]{}
\providecommand*\mciteSetBstMaxWidthForm[2]{}
\providecommand*\mciteBstWouldAddEndPuncttrue
  {\def\EndOfBibitem{\unskip.}}
\providecommand*\mciteBstWouldAddEndPunctfalse
  {\let\EndOfBibitem\relax}
\providecommand*\mciteSetBstMidEndSepPunct[3]{}
\providecommand*\mciteSetBstSublistLabelBeginEnd[3]{}
\providecommand*\EndOfBibitem{}
\mciteSetBstSublistMode{f}
\mciteSetBstMaxWidthForm{subitem}{(\alph{mcitesubitemcount})}
\mciteSetBstSublistLabelBeginEnd
  {\mcitemaxwidthsubitemform\space}
  {\relax}
  {\relax}

\bibitem[Fei \latin{et~al.}(2018)Fei, Zhao, Palomaki, Sun, Miller, Zhao, Yan,
  Xu, and Cobden]{Fei2018}
Fei,~Z.; Zhao,~W.; Palomaki,~T.~A.; Sun,~B.; Miller,~M.~K.; Zhao,~Z.; Yan,~J.;
  Xu,~X.; Cobden,~D.~H. Ferroelectric switching of a two-dimensional metal.
  \emph{Nature} \textbf{2018}, \emph{560}, 336--339\relax
\mciteBstWouldAddEndPuncttrue
\mciteSetBstMidEndSepPunct{\mcitedefaultmidpunct}
{\mcitedefaultendpunct}{\mcitedefaultseppunct}\relax
\EndOfBibitem
\bibitem[Wu and Li(2021)Wu, and Li]{Wu2021}
Wu,~M.; Li,~J. Sliding ferroelectricity in {2D} van der {Waals} materials:
  related physics and future opportunities. \emph{Proc. Natl. Acad. Sci.
  U.S.A.} \textbf{2021}, \emph{118}, e2115703118\relax
\mciteBstWouldAddEndPuncttrue
\mciteSetBstMidEndSepPunct{\mcitedefaultmidpunct}
{\mcitedefaultendpunct}{\mcitedefaultseppunct}\relax
\EndOfBibitem
\bibitem[Vizner~Stern \latin{et~al.}(2021)Vizner~Stern, Waschitz, Cao, Nevo,
  Watanabe, Taniguchi, Sela, Urbakh, Hod, and Ben~Shalom]{Vizner2021}
Vizner~Stern,~M.; Waschitz,~Y.; Cao,~W.; Nevo,~I.; Watanabe,~K.; Taniguchi,~T.;
  Sela,~E.; Urbakh,~M.; Hod,~O.; Ben~Shalom,~M. Interfacial ferroelectricity by
  van der {Waals} sliding. \emph{Science} \textbf{2021}, \emph{372},
  1462--1466\relax
\mciteBstWouldAddEndPuncttrue
\mciteSetBstMidEndSepPunct{\mcitedefaultmidpunct}
{\mcitedefaultendpunct}{\mcitedefaultseppunct}\relax
\EndOfBibitem
\bibitem[Yasuda \latin{et~al.}(2021)Yasuda, Wang, Watanabe, Taniguchi, and
  Jarillo-Herrero]{Yasuda2021}
Yasuda,~K.; Wang,~X.; Watanabe,~K.; Taniguchi,~T.; Jarillo-Herrero,~P.
  Stacking-engineered ferroelectricity in bilayer boron nitride. \emph{Science}
  \textbf{2021}, \emph{372}, 1458--1462\relax
\mciteBstWouldAddEndPuncttrue
\mciteSetBstMidEndSepPunct{\mcitedefaultmidpunct}
{\mcitedefaultendpunct}{\mcitedefaultseppunct}\relax
\EndOfBibitem
\bibitem[Wan \latin{et~al.}(2022)Wan, Hu, Mao, Fu, Yuan, Song, Gan, Xu, Xue,
  Cheng, Huang, Yang, Dai, Zeng, and Kan]{Wan2022}
Wan,~Y.; Hu,~T.; Mao,~X.; Fu,~J.; Yuan,~K.; Song,~Y.; Gan,~X.; Xu,~X.; Xue,~M.;
  Cheng,~X.; Huang,~C.; Yang,~J.; Dai,~L.; Zeng,~H.; Kan,~E. Room-temperature
  ferroelectricity in 1{T}$^{\prime}$-{ReS}$_2$ multilayers. \emph{Phys. Rev.
  Lett.} \textbf{2022}, \emph{128}, 067601\relax
\mciteBstWouldAddEndPuncttrue
\mciteSetBstMidEndSepPunct{\mcitedefaultmidpunct}
{\mcitedefaultendpunct}{\mcitedefaultseppunct}\relax
\EndOfBibitem
\bibitem[Wang \latin{et~al.}(2022)Wang, Yasuda, Zhang, Liu, Watanabe,
  Taniguchi, Hone, Fu, and Jarillo-Herrero]{Wang2022}
Wang,~X.; Yasuda,~K.; Zhang,~Y.; Liu,~S.; Watanabe,~K.; Taniguchi,~T.;
  Hone,~J.; Fu,~L.; Jarillo-Herrero,~P. Interfacial ferroelectricity in
  rhombohedral-stacked bilayer transition metal dichalcogenides. \emph{Nat.
  Nanotechnol.} \textbf{2022}, \emph{17}, 367--371\relax
\mciteBstWouldAddEndPuncttrue
\mciteSetBstMidEndSepPunct{\mcitedefaultmidpunct}
{\mcitedefaultendpunct}{\mcitedefaultseppunct}\relax
\EndOfBibitem
\bibitem[Weston \latin{et~al.}(2022)Weston, Castanon, Enaldiev, Ferreira,
  Bhattacharjee, Xu, Corte-León, Wu, Clark, Summerfield, Hashimoto, Gao, Wang,
  Hamer, Read, Fumagalli, Kretinin, Haigh, Kazakova, Geim, Fal’ko, and
  Gorbachev]{Weston2022}
Weston,~A. \latin{et~al.}  Interfacial ferroelectricity in marginally twisted
  {2D} semiconductors. \emph{Nat. Nanotechnol.} \textbf{2022}, \emph{17},
  390--395\relax
\mciteBstWouldAddEndPuncttrue
\mciteSetBstMidEndSepPunct{\mcitedefaultmidpunct}
{\mcitedefaultendpunct}{\mcitedefaultseppunct}\relax
\EndOfBibitem
\bibitem[Chang \latin{et~al.}(2016)Chang, Liu, Lin, Wang, Zhao, Zhang, Jin,
  Zhong, Hu, Duan, Zhang, Fu, Xue, Chen, and Ji]{Chang2016}
Chang,~K.; Liu,~J.; Lin,~H.; Wang,~N.; Zhao,~K.; Zhang,~A.; Jin,~F.; Zhong,~Y.;
  Hu,~X.; Duan,~W.; Zhang,~Q.; Fu,~L.; Xue,~Q.-K.; Chen,~X.; Ji,~S.-H.
  Discovery of robust in-plane ferroelectricity in atomic-thick {SnTe}.
  \emph{Science} \textbf{2016}, \emph{353}, 274--278\relax
\mciteBstWouldAddEndPuncttrue
\mciteSetBstMidEndSepPunct{\mcitedefaultmidpunct}
{\mcitedefaultendpunct}{\mcitedefaultseppunct}\relax
\EndOfBibitem
\bibitem[Liu \latin{et~al.}(2016)Liu, You, Seyler, Li, Yu, Lin, Wang, Zhou,
  Wang, He, Pantelides, Zhou, Sharma, Xu, Ajayan, Wang, and Liu]{Liu2016}
Liu,~F. \latin{et~al.}  Room-temperature ferroelectricity in {CuInP}$_2${S}$_6$
  ultrathin flakes. \emph{Nat. Commun.} \textbf{2016}, \emph{7}, 12357\relax
\mciteBstWouldAddEndPuncttrue
\mciteSetBstMidEndSepPunct{\mcitedefaultmidpunct}
{\mcitedefaultendpunct}{\mcitedefaultseppunct}\relax
\EndOfBibitem
\bibitem[Xiao \latin{et~al.}(2018)Xiao, Zhu, Wang, Feng, Hu, Dasgupta, Han,
  Wang, Muller, Martin, Hu, and Zhang]{Xiao2018}
Xiao,~J.; Zhu,~H.; Wang,~Y.; Feng,~W.; Hu,~Y.; Dasgupta,~A.; Han,~Y.; Wang,~Y.;
  Muller,~D.~A.; Martin,~L.~W.; Hu,~P.; Zhang,~X. Intrinsic two-dimensional
  ferroelectricity with dipole locking. \emph{Phys. Rev. Lett.} \textbf{2018},
  \emph{120}, 227601\relax
\mciteBstWouldAddEndPuncttrue
\mciteSetBstMidEndSepPunct{\mcitedefaultmidpunct}
{\mcitedefaultendpunct}{\mcitedefaultseppunct}\relax
\EndOfBibitem
\bibitem[Ke \latin{et~al.}(2025)Ke, Liu, and Liu]{Ke2025}
Ke,~C.; Liu,~F.; Liu,~S. Superlubric motion of wavelike domain walls in sliding
  ferroelectrics. \emph{Phys. Rev. Lett.} \textbf{2025}, \emph{135},
  046201\relax
\mciteBstWouldAddEndPuncttrue
\mciteSetBstMidEndSepPunct{\mcitedefaultmidpunct}
{\mcitedefaultendpunct}{\mcitedefaultseppunct}\relax
\EndOfBibitem
\bibitem[Wang \latin{et~al.}(2021)Wang, Zhu, Deng, Duan, Chen, Zeng, Zhou, Fu,
  You, Liu, Edgar, Yu, and Liu]{Wang2021}
Wang,~X.; Zhu,~C.; Deng,~Y.; Duan,~R.; Chen,~J.; Zeng,~Q.; Zhou,~J.; Fu,~Q.;
  You,~L.; Liu,~S.; Edgar,~J.~H.; Yu,~P.; Liu,~Z. Van der {Waals} engineering
  of ferroelectric heterostructures for long-retention memory. \emph{Nat.
  Commun.} \textbf{2021}, \emph{12}, 1109\relax
\mciteBstWouldAddEndPuncttrue
\mciteSetBstMidEndSepPunct{\mcitedefaultmidpunct}
{\mcitedefaultendpunct}{\mcitedefaultseppunct}\relax
\EndOfBibitem
\bibitem[Li \latin{et~al.}(2024)Li, Qin, Wang, Xi, Huang, Zhao, Peng, Chen,
  Pan, Zhu, Cui, Yang, Yang, Meng, Shi, Bai, Liu, Li, Tang, Liu, Du, and
  Zhang]{Li2024}
Li,~X. \latin{et~al.}  Sliding ferroelectric memories and synapses based on
  rhombohedral-stacked bilayer {MoS}$_2$. \emph{Nat. Commun.} \textbf{2024},
  \emph{15}, 10921\relax
\mciteBstWouldAddEndPuncttrue
\mciteSetBstMidEndSepPunct{\mcitedefaultmidpunct}
{\mcitedefaultendpunct}{\mcitedefaultseppunct}\relax
\EndOfBibitem
\bibitem[Wu \latin{et~al.}(2023)Wu, Zhang, Feng, Wang, Zhou, Zeng, Dong, Zhu,
  Yang, Zhao, Gong, Zhang, Tian, Duan, Liu, Wang, Chu, and Liu]{Wu2023}
Wu,~G. \latin{et~al.}  Ferroelectric-defined reconfigurable homojunctions for
  in-memory sensing and computing. \emph{Nat. Mater.} \textbf{2023}, \emph{22},
  1499--1506\relax
\mciteBstWouldAddEndPuncttrue
\mciteSetBstMidEndSepPunct{\mcitedefaultmidpunct}
{\mcitedefaultendpunct}{\mcitedefaultseppunct}\relax
\EndOfBibitem
\bibitem[Bian \latin{et~al.}(2024)Bian, He, Pan, Li, Cao, Meng, Chen, Liu,
  Zhong, Li, and Liu]{Bian2024}
Bian,~R.; He,~R.; Pan,~E.; Li,~Z.; Cao,~G.; Meng,~P.; Chen,~J.; Liu,~Q.;
  Zhong,~Z.; Li,~W.; Liu,~F. Developing fatigue-resistant ferroelectrics using
  interlayer sliding switching. \emph{Science} \textbf{2024}, \emph{385},
  57--62\relax
\mciteBstWouldAddEndPuncttrue
\mciteSetBstMidEndSepPunct{\mcitedefaultmidpunct}
{\mcitedefaultendpunct}{\mcitedefaultseppunct}\relax
\EndOfBibitem
\bibitem[Yasuda \latin{et~al.}(2024)Yasuda, Zalys-Geller, Wang, Bennett,
  Cheema, Watanabe, Taniguchi, Kaxiras, Jarillo-Herrero, and
  Ashoori]{Yasuda2024}
Yasuda,~K.; Zalys-Geller,~E.; Wang,~X.; Bennett,~D.; Cheema,~S.~S.;
  Watanabe,~K.; Taniguchi,~T.; Kaxiras,~E.; Jarillo-Herrero,~P.; Ashoori,~R.
  Ultrafast high-endurance memory based on sliding ferroelectrics.
  \emph{Science} \textbf{2024}, \emph{385}, 53--56\relax
\mciteBstWouldAddEndPuncttrue
\mciteSetBstMidEndSepPunct{\mcitedefaultmidpunct}
{\mcitedefaultendpunct}{\mcitedefaultseppunct}\relax
\EndOfBibitem
\bibitem[Chen \latin{et~al.}(2024)Chen, Xie, Cheng, Yang, Li, Chen, Li, Xie,
  Watanabe, Taniguchi, He, Wu, Liang, and Miao]{Chen2024}
Chen,~M.; Xie,~Y.; Cheng,~B.; Yang,~Z.; Li,~X.-Z.; Chen,~F.; Li,~Q.; Xie,~J.;
  Watanabe,~K.; Taniguchi,~T.; He,~W.-Y.; Wu,~M.; Liang,~S.-J.; Miao,~F.
  Selective and quasi-continuous switching of ferroelectric {Chern} insulator
  devices for neuromorphic computing. \emph{Nat. Nanotechnol.} \textbf{2024},
  \emph{19}, 962--969\relax
\mciteBstWouldAddEndPuncttrue
\mciteSetBstMidEndSepPunct{\mcitedefaultmidpunct}
{\mcitedefaultendpunct}{\mcitedefaultseppunct}\relax
\EndOfBibitem
\bibitem[Xiao \latin{et~al.}(2020)Xiao, Wang, Wang, Pemmaraju, Wang, Muscher,
  Sie, Nyby, Devereaux, Qian, Zhang, and Lindenberg]{Xiao2020}
Xiao,~J.; Wang,~Y.; Wang,~H.; Pemmaraju,~C.~D.; Wang,~S.; Muscher,~P.;
  Sie,~E.~J.; Nyby,~C.~M.; Devereaux,~T.~P.; Qian,~X.; Zhang,~X.;
  Lindenberg,~A.~M. Berry curvature memory through electrically driven stacking
  transitions. \emph{Nat. Phys.} \textbf{2020}, \emph{16}, 1028--1034\relax
\mciteBstWouldAddEndPuncttrue
\mciteSetBstMidEndSepPunct{\mcitedefaultmidpunct}
{\mcitedefaultendpunct}{\mcitedefaultseppunct}\relax
\EndOfBibitem
\bibitem[Jindal \latin{et~al.}(2023)Jindal, Saha, Li, Taniguchi, Watanabe,
  Hone, Birol, Fernandes, Dean, Pasupathy, and Rhodes]{Jindal2023}
Jindal,~A.; Saha,~A.; Li,~Z.~Z.; Taniguchi,~T.; Watanabe,~K.; Hone,~J.~C.;
  Birol,~T.; Fernandes,~R.~M.; Dean,~C.~R.; Pasupathy,~A.~N.; Rhodes,~D.~A.
  Coupled ferroelectricity and superconductivity in bilayer
  {T$_{\mathrm{d}}$-MoTe$_2$}. \emph{Nature} \textbf{2023}, \emph{613},
  48--52\relax
\mciteBstWouldAddEndPuncttrue
\mciteSetBstMidEndSepPunct{\mcitedefaultmidpunct}
{\mcitedefaultendpunct}{\mcitedefaultseppunct}\relax
\EndOfBibitem
\bibitem[Klein \latin{et~al.}(2023)Klein, Xia, MacNeill, Watanabe, Taniguchi,
  and Jarillo-Herrero]{Klein2023}
Klein,~D.~R.; Xia,~L.-Q.; MacNeill,~D.; Watanabe,~K.; Taniguchi,~T.;
  Jarillo-Herrero,~P. Electrical switching of a bistable moiré superconductor.
  \emph{Nat. Nanotechnol.} \textbf{2023}, \emph{18}, 331--335\relax
\mciteBstWouldAddEndPuncttrue
\mciteSetBstMidEndSepPunct{\mcitedefaultmidpunct}
{\mcitedefaultendpunct}{\mcitedefaultseppunct}\relax
\EndOfBibitem
\bibitem[Mak \latin{et~al.}(2010)Mak, Lee, Hone, Shan, and Heinz]{Mak2010}
Mak,~K.~F.; Lee,~C.; Hone,~J.; Shan,~J.; Heinz,~T.~F. Atomically thin
  {MoS}$_2$: a new direct-gap semiconductor. \emph{Phys. Rev. Lett.}
  \textbf{2010}, \emph{105}, 136805\relax
\mciteBstWouldAddEndPuncttrue
\mciteSetBstMidEndSepPunct{\mcitedefaultmidpunct}
{\mcitedefaultendpunct}{\mcitedefaultseppunct}\relax
\EndOfBibitem
\bibitem[Splendiani \latin{et~al.}(2010)Splendiani, Sun, Zhang, Li, Kim, Chim,
  Galli, and Wang]{Splendiani2010}
Splendiani,~A.; Sun,~L.; Zhang,~Y.; Li,~T.; Kim,~J.; Chim,~C.-Y.; Galli,~G.;
  Wang,~F. Emerging photoluminescence in monolayer {MoS}$_2$. \emph{Nano Lett.}
  \textbf{2010}, \emph{10}, 1271--1275\relax
\mciteBstWouldAddEndPuncttrue
\mciteSetBstMidEndSepPunct{\mcitedefaultmidpunct}
{\mcitedefaultendpunct}{\mcitedefaultseppunct}\relax
\EndOfBibitem
\bibitem[Zhao \latin{et~al.}(2013)Zhao, Ghorannevis, Chu, Toh, Kloc, Tan, and
  Eda]{Zhao2013}
Zhao,~W.; Ghorannevis,~Z.; Chu,~L.; Toh,~M.; Kloc,~C.; Tan,~P.-H.; Eda,~G.
  Evolution of electronic structure in atomically thin sheets of {WS}$_2$ and
  {WSe}$_2$. \emph{ACS Nano} \textbf{2013}, \emph{7}, 791--797\relax
\mciteBstWouldAddEndPuncttrue
\mciteSetBstMidEndSepPunct{\mcitedefaultmidpunct}
{\mcitedefaultendpunct}{\mcitedefaultseppunct}\relax
\EndOfBibitem
\bibitem[Mak and Shan(2016)Mak, and Shan]{Mak2016}
Mak,~K.~F.; Shan,~J. Photonics and optoelectronics of {2D} semiconductor
  transition metal dichalcogenides. \emph{Nat. Photonics} \textbf{2016},
  \emph{10}, 216--226\relax
\mciteBstWouldAddEndPuncttrue
\mciteSetBstMidEndSepPunct{\mcitedefaultmidpunct}
{\mcitedefaultendpunct}{\mcitedefaultseppunct}\relax
\EndOfBibitem
\bibitem[Wang \latin{et~al.}(2018)Wang, Chernikov, Glazov, Heinz, Marie, Amand,
  and Urbaszek]{Wang2018RMP}
Wang,~G.; Chernikov,~A.; Glazov,~M.~M.; Heinz,~T.~F.; Marie,~X.; Amand,~T.;
  Urbaszek,~B. Colloquium: excitons in atomically thin transition metal
  dichalcogenides. \emph{Rev. Mod. Phys.} \textbf{2018}, \emph{90},
  021001\relax
\mciteBstWouldAddEndPuncttrue
\mciteSetBstMidEndSepPunct{\mcitedefaultmidpunct}
{\mcitedefaultendpunct}{\mcitedefaultseppunct}\relax
\EndOfBibitem
\bibitem[Rivera \latin{et~al.}(2015)Rivera, Schaibley, Jones, Ross, Wu,
  Aivazian, Klement, Seyler, Clark, Ghimire, Yan, Mandrus, Yao, and
  Xu]{Rivera2015}
Rivera,~P.; Schaibley,~J.~R.; Jones,~A.~M.; Ross,~J.~S.; Wu,~S.; Aivazian,~G.;
  Klement,~P.; Seyler,~K.; Clark,~G.; Ghimire,~N.~J.; Yan,~J.; Mandrus,~D.~G.;
  Yao,~W.; Xu,~X. Observation of long-lived interlayer excitons in monolayer
  {MoSe}$_2$-{WSe}$_2$ heterostructures. \emph{Nat. Commun.} \textbf{2015},
  \emph{6}, 6242\relax
\mciteBstWouldAddEndPuncttrue
\mciteSetBstMidEndSepPunct{\mcitedefaultmidpunct}
{\mcitedefaultendpunct}{\mcitedefaultseppunct}\relax
\EndOfBibitem
\bibitem[Wang \latin{et~al.}(2018)Wang, Chiu, Honz, Mak, and Shan]{Wang2018}
Wang,~Z.; Chiu,~Y.-H.; Honz,~K.; Mak,~K.~F.; Shan,~J. Electrical tuning of
  interlayer exciton gases in $\mathrm{WSe}_2$ bilayers. \emph{Nano Lett.}
  \textbf{2018}, \emph{18}, 137--143\relax
\mciteBstWouldAddEndPuncttrue
\mciteSetBstMidEndSepPunct{\mcitedefaultmidpunct}
{\mcitedefaultendpunct}{\mcitedefaultseppunct}\relax
\EndOfBibitem
\bibitem[Peimyoo \latin{et~al.}(2021)Peimyoo, Deilmann, Withers, Escolar,
  Nutting, Taniguchi, Watanabe, Taghizadeh, Craciun, Thygesen, and
  Russo]{Peimyoo2021}
Peimyoo,~N.; Deilmann,~T.; Withers,~F.; Escolar,~J.; Nutting,~D.;
  Taniguchi,~T.; Watanabe,~K.; Taghizadeh,~A.; Craciun,~M.~F.; Thygesen,~K.~S.;
  Russo,~S. Electrical tuning of optically active interlayer excitons in
  bilayer $\mathrm{MoS}_2$. \emph{Nat. Nanotechnol.} \textbf{2021}, \emph{16},
  888--893\relax
\mciteBstWouldAddEndPuncttrue
\mciteSetBstMidEndSepPunct{\mcitedefaultmidpunct}
{\mcitedefaultendpunct}{\mcitedefaultseppunct}\relax
\EndOfBibitem
\bibitem[Huang \latin{et~al.}(2022)Huang, Zhao, Bo, Chu, Tian, Liu, Yuan, Wu,
  Zhao, Xian, Watanabe, Taniguchi, Yang, Shi, Du, Sun, Meng, Yang, and
  Zhang]{Huang2022}
Huang,~Z. \latin{et~al.}  Spatially indirect intervalley excitons in bilayer
  {WSe$_2$}. \emph{Phys. Rev. B} \textbf{2022}, \emph{105}, L041409\relax
\mciteBstWouldAddEndPuncttrue
\mciteSetBstMidEndSepPunct{\mcitedefaultmidpunct}
{\mcitedefaultendpunct}{\mcitedefaultseppunct}\relax
\EndOfBibitem
\bibitem[Feng \latin{et~al.}(2024)Feng, Campbell, Brotons-Gisbert,
  Andres-Penares, Baek, Taniguchi, Watanabe, Urbaszek, Gerber, and
  Gerardot]{Feng2024}
Feng,~S.; Campbell,~A.~J.; Brotons-Gisbert,~M.; Andres-Penares,~D.; Baek,~H.;
  Taniguchi,~T.; Watanabe,~K.; Urbaszek,~B.; Gerber,~I.~C.; Gerardot,~B.~D.
  Highly tunable ground and excited state excitonic dipoles in multilayer
  {2H}-{MoSe}$_2$. \emph{Nat. Commun.} \textbf{2024}, \emph{15}, 4377\relax
\mciteBstWouldAddEndPuncttrue
\mciteSetBstMidEndSepPunct{\mcitedefaultmidpunct}
{\mcitedefaultendpunct}{\mcitedefaultseppunct}\relax
\EndOfBibitem
\bibitem[Sung \latin{et~al.}(2020)Sung, Zhou, Scuri, Z{\'o}lyomi, Andersen,
  Yoo, Wild, Joe, Gelly, Heo, Magorrian, B{\'e}rub{\'e}, Mier~Valdivia,
  Taniguchi, Watanabe, Lukin, Kim, Fal'ko, and Park]{Sung2020}
Sung,~J. \latin{et~al.}  Broken mirror symmetry in excitonic response of
  reconstructed domains in twisted $\mathrm{MoSe}_2/\mathrm{MoSe}_2$ bilayers.
  \emph{Nat. Nanotechnol.} \textbf{2020}, \emph{15}, 750--754\relax
\mciteBstWouldAddEndPuncttrue
\mciteSetBstMidEndSepPunct{\mcitedefaultmidpunct}
{\mcitedefaultendpunct}{\mcitedefaultseppunct}\relax
\EndOfBibitem
\bibitem[Andersen \latin{et~al.}(2021)Andersen, Scuri, Sushko, De~Greve, Sung,
  Zhou, Wild, Gelly, Heo, Bérubé, Joe, Jauregui, Watanabe, Taniguchi, Kim,
  Park, and Lukin]{Andersen2021}
Andersen,~T.~I. \latin{et~al.}  Excitons in a reconstructed moiré potential in
  twisted $\mathrm{WSe}_2$/$\mathrm{WSe}_2$ homobilayers. \emph{Nat. Mater.}
  \textbf{2021}, \emph{20}, 480--487\relax
\mciteBstWouldAddEndPuncttrue
\mciteSetBstMidEndSepPunct{\mcitedefaultmidpunct}
{\mcitedefaultendpunct}{\mcitedefaultseppunct}\relax
\EndOfBibitem
\bibitem[Liang \latin{et~al.}(2022)Liang, Yang, Wu, Dadap, Watanabe, Taniguchi,
  and Ye]{Liang2022}
Liang,~J.; Yang,~D.; Wu,~J.; Dadap,~J.~I.; Watanabe,~K.; Taniguchi,~T.; Ye,~Z.
  Optically probing the asymmetric interlayer coupling in rhombohedral-stacked
  {MoS}$_2$ bilayer. \emph{Phys. Rev. X} \textbf{2022}, \emph{12}, 041005\relax
\mciteBstWouldAddEndPuncttrue
\mciteSetBstMidEndSepPunct{\mcitedefaultmidpunct}
{\mcitedefaultendpunct}{\mcitedefaultseppunct}\relax
\EndOfBibitem
\bibitem[Yang \latin{et~al.}(2024)Yang, Liang, Wu, Xiao, Dadap, Watanabe,
  Taniguchi, and Ye]{Yang2024}
Yang,~D.; Liang,~J.; Wu,~J.; Xiao,~Y.; Dadap,~J.~I.; Watanabe,~K.;
  Taniguchi,~T.; Ye,~Z. Non-volatile electrical polarization switching via
  domain wall release in {3R}-{MoS}$_2$ bilayer. \emph{Nat. Commun.}
  \textbf{2024}, \emph{15}, 1389\relax
\mciteBstWouldAddEndPuncttrue
\mciteSetBstMidEndSepPunct{\mcitedefaultmidpunct}
{\mcitedefaultendpunct}{\mcitedefaultseppunct}\relax
\EndOfBibitem
\bibitem[Deb \latin{et~al.}(2024)Deb, Krause, Faria~Junior, Kempf, Schwartz,
  Watanabe, Taniguchi, Fabian, and Korn]{Deb2024}
Deb,~S.; Krause,~J.; Faria~Junior,~P.~E.; Kempf,~M.~A.; Schwartz,~R.;
  Watanabe,~K.; Taniguchi,~T.; Fabian,~J.; Korn,~T. Excitonic signatures of
  ferroelectric order in parallel-stacked {MoS}$_2$. \emph{Nat. Commun.}
  \textbf{2024}, \emph{15}, 7595\relax
\mciteBstWouldAddEndPuncttrue
\mciteSetBstMidEndSepPunct{\mcitedefaultmidpunct}
{\mcitedefaultendpunct}{\mcitedefaultseppunct}\relax
\EndOfBibitem
\bibitem[Liang \latin{et~al.}(2025)Liang, Xie, Yang, Guo, Watanabe, Taniguchi,
  Dadap, Jones, and Ye]{Liang2025PRX}
Liang,~J.; Xie,~Y.; Yang,~D.; Guo,~S.; Watanabe,~K.; Taniguchi,~T.;
  Dadap,~J.~I.; Jones,~D.; Ye,~Z. Nanosecond ferroelectric switching of
  intralayer excitons in bilayer {3R}-{MoS}$_2$ through {Coulomb} engineering.
  \emph{Phys. Rev. X} \textbf{2025}, \emph{15}, 021081\relax
\mciteBstWouldAddEndPuncttrue
\mciteSetBstMidEndSepPunct{\mcitedefaultmidpunct}
{\mcitedefaultendpunct}{\mcitedefaultseppunct}\relax
\EndOfBibitem
\bibitem[Liang \latin{et~al.}(2025)Liang, Yang, Wu, Xiao, Watanabe, Taniguchi,
  Dadap, and Ye]{Liang2025}
Liang,~J.; Yang,~D.; Wu,~J.; Xiao,~Y.; Watanabe,~K.; Taniguchi,~T.;
  Dadap,~J.~I.; Ye,~Z. Resolving polarization switching pathways of sliding
  ferroelectricity in trilayer {3R}-{MoS$_2$}. \emph{Nat. Nanotechnol.}
  \textbf{2025}, \emph{20}, 500--506\relax
\mciteBstWouldAddEndPuncttrue
\mciteSetBstMidEndSepPunct{\mcitedefaultmidpunct}
{\mcitedefaultendpunct}{\mcitedefaultseppunct}\relax
\EndOfBibitem
\bibitem[Ouyang \latin{et~al.}(2025)Ouyang, Cha, Sun, Taniguchi, Watanabe,
  Gabor, and Lui]{Ouyang2025}
Ouyang,~T.; Cha,~S.; Sun,~Y.; Taniguchi,~T.; Watanabe,~K.; Gabor,~N.~M.;
  Lui,~C.~H. Electrically switching ferroelectric order in {3R}-{MoS}$_2$
  layers. \emph{Nano Lett.} \textbf{2025}, \emph{25}, 1459--1465\relax
\mciteBstWouldAddEndPuncttrue
\mciteSetBstMidEndSepPunct{\mcitedefaultmidpunct}
{\mcitedefaultendpunct}{\mcitedefaultseppunct}\relax
\EndOfBibitem
\bibitem[Li \latin{et~al.}(2026)Li, Thor, Kourmoulakis, Ivanova, Taniguchi,
  Watanabe, Yu, Brotons-Gisbert, and Gerardot]{Li2026}
Li,~Z.; Thor,~P.; Kourmoulakis,~G.; Ivanova,~T.~V.; Taniguchi,~T.;
  Watanabe,~K.; Yu,~H.; Brotons-Gisbert,~M.; Gerardot,~B.~D. Highly tunable
  band structure in ferroelectric {R}-stacked bilayer $\mathrm{WSe}_2$.
  \emph{Nat. Commun.} \textbf{2026}, \emph{17}, 2457\relax
\mciteBstWouldAddEndPuncttrue
\mciteSetBstMidEndSepPunct{\mcitedefaultmidpunct}
{\mcitedefaultendpunct}{\mcitedefaultseppunct}\relax
\EndOfBibitem
\bibitem[Li and Wu(2017)Li, and Wu]{Li2017}
Li,~L.; Wu,~M. Binary compound bilayer and multilayer with vertical
  polarizations: two-dimensional ferroelectrics, multiferroics, and
  nanogenerators. \emph{ACS Nano} \textbf{2017}, \emph{11}, 6382--6388\relax
\mciteBstWouldAddEndPuncttrue
\mciteSetBstMidEndSepPunct{\mcitedefaultmidpunct}
{\mcitedefaultendpunct}{\mcitedefaultseppunct}\relax
\EndOfBibitem
\bibitem[Naik and Jain(2018)Naik, and Jain]{Naik2018}
Naik,~M.~H.; Jain,~M. Ultraflatbands and shear solitons in moir\'e patterns of
  twisted bilayer transition metal dichalcogenides. \emph{Phys. Rev. Lett.}
  \textbf{2018}, \emph{121}, 266401\relax
\mciteBstWouldAddEndPuncttrue
\mciteSetBstMidEndSepPunct{\mcitedefaultmidpunct}
{\mcitedefaultendpunct}{\mcitedefaultseppunct}\relax
\EndOfBibitem
\bibitem[Seyler \latin{et~al.}(2015)Seyler, Schaibley, Gong, Rivera, Jones, Wu,
  Yan, Mandrus, Yao, and Xu]{Seyler2015}
Seyler,~K.~L.; Schaibley,~J.~R.; Gong,~P.; Rivera,~P.; Jones,~A.~M.; Wu,~S.;
  Yan,~J.; Mandrus,~D.~G.; Yao,~W.; Xu,~X. Electrical control of
  second-harmonic generation in a $\mathrm{WSe}_2$ monolayer transistor.
  \emph{Nat. Nanotechnol.} \textbf{2015}, \emph{10}, 407--411\relax
\mciteBstWouldAddEndPuncttrue
\mciteSetBstMidEndSepPunct{\mcitedefaultmidpunct}
{\mcitedefaultendpunct}{\mcitedefaultseppunct}\relax
\EndOfBibitem
\bibitem[Scuri \latin{et~al.}(2020)Scuri, Andersen, Zhou, Wild, Sung, Gelly,
  B{\'e}rub{\'e}, Heo, Shao, Joe, Mier~Valdivia, Taniguchi, Watanabe, Lon{\v
  c}ar, Kim, Lukin, and Park]{Scuri2020}
Scuri,~G. \latin{et~al.}  Electrically tunable valley dynamics in twisted
  {WSe}$_2$/{WSe}$_2$ bilayers. \emph{Phys. Rev. Lett.} \textbf{2020},
  \emph{124}, 217403\relax
\mciteBstWouldAddEndPuncttrue
\mciteSetBstMidEndSepPunct{\mcitedefaultmidpunct}
{\mcitedefaultendpunct}{\mcitedefaultseppunct}\relax
\EndOfBibitem
\bibitem[Li \latin{et~al.}(2022)Li, F\"orste, Watanabe, Taniguchi, Urbaszek,
  Baimuratov, Gerber, H\"ogele, and Bilgin]{Li2022}
Li,~Z.; F\"orste,~J.; Watanabe,~K.; Taniguchi,~T.; Urbaszek,~B.;
  Baimuratov,~A.~S.; Gerber,~I.~C.; H\"ogele,~A.; Bilgin,~I. Stacking-dependent
  exciton multiplicity in {WSe$_2$} bilayers. \emph{Phys. Rev. B}
  \textbf{2022}, \emph{106}, 045411\relax
\mciteBstWouldAddEndPuncttrue
\mciteSetBstMidEndSepPunct{\mcitedefaultmidpunct}
{\mcitedefaultendpunct}{\mcitedefaultseppunct}\relax
\EndOfBibitem
\bibitem[Altaiary \latin{et~al.}(2022)Altaiary, Liu, Liang, Hsiao, van Baren,
  Taniguchi, Watanabe, Gabor, Chang, and Lui]{Altaiary2022}
Altaiary,~M.~M.; Liu,~E.; Liang,~C.-T.; Hsiao,~F.-C.; van Baren,~J.;
  Taniguchi,~T.; Watanabe,~K.; Gabor,~N.~M.; Chang,~Y.-C.; Lui,~C.~H.
  Electrically switchable intervalley excitons with strong two-phonon
  scattering in bilayer {WSe$_2$}. \emph{Nano Lett.} \textbf{2022}, \emph{22},
  1829--1835\relax
\mciteBstWouldAddEndPuncttrue
\mciteSetBstMidEndSepPunct{\mcitedefaultmidpunct}
{\mcitedefaultendpunct}{\mcitedefaultseppunct}\relax
\EndOfBibitem
\bibitem[Ferreira \latin{et~al.}(2021)Ferreira, Enaldiev, Fal’ko, and
  Magorrian]{Ferreira2021}
Ferreira,~F.; Enaldiev,~V.~V.; Fal’ko,~V.~I.; Magorrian,~S.~J. Weak
  ferroelectric charge transfer in layer-asymmetric bilayers of {2D}
  semiconductors. \emph{Sci. Rep.} \textbf{2021}, \emph{11}, 13422\relax
\mciteBstWouldAddEndPuncttrue
\mciteSetBstMidEndSepPunct{\mcitedefaultmidpunct}
{\mcitedefaultendpunct}{\mcitedefaultseppunct}\relax
\EndOfBibitem
\bibitem[Deb \latin{et~al.}(2022)Deb, Cao, Raab, Watanabe, Taniguchi,
  Goldstein, Kronik, Urbakh, Hod, and Ben~Shalom]{Deb2022}
Deb,~S.; Cao,~W.; Raab,~N.; Watanabe,~K.; Taniguchi,~T.; Goldstein,~M.;
  Kronik,~L.; Urbakh,~M.; Hod,~O.; Ben~Shalom,~M. Cumulative polarization in
  conductive interfacial ferroelectrics. \emph{Nature} \textbf{2022},
  \emph{612}, 465--469\relax
\mciteBstWouldAddEndPuncttrue
\mciteSetBstMidEndSepPunct{\mcitedefaultmidpunct}
{\mcitedefaultendpunct}{\mcitedefaultseppunct}\relax
\EndOfBibitem
\bibitem[Tong \latin{et~al.}(2017)Tong, Yu, Zhu, Wang, Xu, and Yao]{Tong2017}
Tong,~Q.; Yu,~H.; Zhu,~Q.; Wang,~Y.; Xu,~X.; Yao,~W. Topological mosaics in
  moiré superlattices of van der {W}aals heterobilayers. \emph{Nat. Phys.}
  \textbf{2017}, \emph{13}, 356--362\relax
\mciteBstWouldAddEndPuncttrue
\mciteSetBstMidEndSepPunct{\mcitedefaultmidpunct}
{\mcitedefaultendpunct}{\mcitedefaultseppunct}\relax
\EndOfBibitem
\bibitem[Lin \latin{et~al.}(2021)Lin, Holler, Bauer, Parzefall, Scheuck, Peng,
  Korn, Bange, Lupton, and Schüller]{Lin2021}
Lin,~K.-Q.; Holler,~J.; Bauer,~J.~M.; Parzefall,~P.; Scheuck,~M.; Peng,~B.;
  Korn,~T.; Bange,~S.; Lupton,~J.~M.; Schüller,~C. Large-scale mapping of
  moir\'e superlattices by hyperspectral {Raman} imaging. \emph{Adv. Mater.}
  \textbf{2021}, \emph{33}, 2008333\relax
\mciteBstWouldAddEndPuncttrue
\mciteSetBstMidEndSepPunct{\mcitedefaultmidpunct}
{\mcitedefaultendpunct}{\mcitedefaultseppunct}\relax
\EndOfBibitem
\bibitem[Pan \latin{et~al.}(2022)Pan, Li, Rahaman, Milekhin, and Zahn]{Pan2022}
Pan,~Y.; Li,~S.; Rahaman,~M.; Milekhin,~I.; Zahn,~D. R.~T. Signature of lattice
  dynamics in twisted {2D} homo/hetero-bilayers. \emph{2D Mater.}
  \textbf{2022}, \emph{9}, 045018\relax
\mciteBstWouldAddEndPuncttrue
\mciteSetBstMidEndSepPunct{\mcitedefaultmidpunct}
{\mcitedefaultendpunct}{\mcitedefaultseppunct}\relax
\EndOfBibitem
\bibitem[Yu and Cardona(2010)Yu, and Cardona]{Yu2010}
Yu,~P.~Y.; Cardona,~M. \emph{Fundamentals of Semiconductors: Physics and
  Materials Properties}, 4th ed.; Springer: Berlin, Heidelberg, 2010; Chapter
  7, pp 345--426\relax
\mciteBstWouldAddEndPuncttrue
\mciteSetBstMidEndSepPunct{\mcitedefaultmidpunct}
{\mcitedefaultendpunct}{\mcitedefaultseppunct}\relax
\EndOfBibitem
\bibitem[McCreary \latin{et~al.}(2022)McCreary, Phillips, Chuang,
  Wickramaratne, Rosenberger, Hellberg, and Jonker]{McCreary2022}
McCreary,~K.~M.; Phillips,~M.; Chuang,~H.-J.; Wickramaratne,~D.;
  Rosenberger,~M.; Hellberg,~C.~S.; Jonker,~B.~T. Stacking-dependent optical
  properties in bilayer $\mathrm{WSe}_2$. \emph{Nanoscale} \textbf{2022},
  \emph{14}, 147--156\relax
\mciteBstWouldAddEndPuncttrue
\mciteSetBstMidEndSepPunct{\mcitedefaultmidpunct}
{\mcitedefaultendpunct}{\mcitedefaultseppunct}\relax
\EndOfBibitem
\end{mcitethebibliography}
\end{document}